\documentclass[aps,pra,reprint,onecolumn,superscriptaddress,amsmath,amssymb]{revtex4-1}

\usepackage{amssymb}
\usepackage{amsmath}
\usepackage{hyperref}
\usepackage{enumerate}
\usepackage{graphicx}
\usepackage{subfigure}

\begin{document}

\title{Effect of quantum noise on deterministic joint remote state preparation of a qubit state via a GHZ channel}

\author{Ming-Ming Wang}
\email{bluess1982@126.com}
\affiliation{School of Computer Science, Xi'an Polytechnic University, Xi'an 710048, China}
\affiliation{Jiangsu Engineering Center of Network Monitoring, Nanjing University of Information Science \& Technology, Nanjing 210044, China}
\author{Zhi-Guo Qu}
\affiliation{Jiangsu Engineering Center of Network Monitoring, Nanjing University of Information Science \& Technology, Nanjing 210044, China}

\date{\today}

\begin{abstract}
Quantum secure communication brings a new direction for information security. As an important component of quantum secure communication, deterministic joint remote state preparation (DJRSP) could securely transmit a quantum state with 100\% success probability. In this paper, we study how the efficiency of DJRSP is affected when qubits involved in the protocol are subjected to noise or decoherence. Taking a GHZ based DJRSP scheme as an example, we study all types of noise usually encountered in real-world implementations of quantum communication protocols, i.e., the bit-flip, phase-flip (phase-damping), depolarizing, and amplitude-damping noise. Our study shows that the fidelity of the output state depends on the phase factor, the amplitude factor and the noise parameter in the bit-flip noise, while the fidelity only depends on the amplitude factor and the noise parameter in the other three types of noise. And the receiver will get different output states depending on the first preparer's measurement result in the amplitude-damping noise. Our results will be helpful for improving quantum secure communication in real implementation.
\end{abstract}

\maketitle

\section{Introduction}
Quantum information and quantum communication have greatly impacted the development direction of modern science. On the one hand, quantum cryptograph, such as quantum key distribution \cite{BB84}, quantum secret sharing (QSS) \cite{HilleryBuzek-99}, quantum data hiding \cite{TerhalDivincenzo-998,QuChen-997}, quantum signature \cite{Wang-503} and quantum authentication \cite{CurtySantos-422} can achieve high-level security than their classical counterparts \cite{Shamir79,XiaWang-995,XiaWang-996,MaZhou-994,GuoWang-992,RenShen-993}. On the other hand, quantum algorithms, such as Grover’s search algorithm \cite{Grover-225}, can solve a certain problem much faster than classical algorithms \cite{XiaWang-989,FuSun-1050,FuRen-1052}.

In quantum world, quantum entanglement is a crucial resource and an amazing application of entanglement is quantum teleportation \cite{BBC-93}, which can securely transmit a quantum state from a preparer to a remote receiver by virtue of pre-shared entangled resource. If the preparer has already known the information of the state, the transmission can be achieved by RSP \cite{Lo-54,Pati-50,BennettDivincenzo-53} with simpler measurement and less classical communication costs. The original RSP scheme only has one preparer who knows all the information of the prepared state. But for highly sensitive and important information, it might not be reliable to let one person hold everything. To solve this potential problem, joint RSP (JRSP) has been proposed \cite{Xia-40}, which involves at least two preparers. Each preparer holds partial information and only if certain preparers work together can the state be remotely prepared, similar to the idea of secret sharing.
However, a serious problem for most of the previous JRSP schemes \cite{Nguyen-41,Hou-48,Luo-283,Nguyen-283} is that they are probabilistic, i.e., the success probability is less than 1. Recently, a new direction of JRSP, namely deterministic JRSP (DJRSP) has been put forward. Xiao et al. \cite{XiaoLiu-386} introduced the three-step strategy to increase the success probability of JRSP. By adding some classical communication and local operations, the success probability of preparation can be increased to 1. Nguyen et al. \cite{NguyenCao-385} presented two DJRSP schemes of general one- and two-qubit states by using EPR pairs.  Chen et al. \cite{ChenXia-384} extended this idea to realize a DJRSP of an arbitrary three-qubit state by using six EPR pairs. In 2014, we proposed a deterministic JRSP scheme of an arbitrary two-qubit state based on the six-qubit cluster state \cite{Wang-471}.

Quantum noise is an unavoidable factor in practical quantum communication system, which will severely affect the security and reliability of the system \cite{WangWang-1001}. For a RSP scheme, the entanglement shared among participants will turn a pure state into a mixed one in the presence of noise. In recent years, some RSP schemes in noisy environment have been studied. Xiang et al. \cite{XiangLi-941} presented a RSP protocol for mixed state in depolarizing and dephasing channel. Chen et al. \cite{Ai-XiLi-942} investigated remote preparation of an entangled state through a mixed state channel in nonideal conditions. Guan et al. \cite{GuanChen-939} studied a JRSP of an arbitrary two-qubit state in the amplitude-damping and the phase-damping noisy environment. Liang et al. \cite{LiangLiu-983,LiangLiu-1054} investigated a JRSP of a qubit state in different noises by solving Lindblad master equation.
Sharma et al. \cite{SharmaShukla-1057} investigated the effect of amplitude-damping and phase-damping noise on a bidirectional RSP protocol.
Li et al. \cite{LiLiu-1056} investigated a DJRSP of an arbitrary two-qubit state via four EPR pairs channel which are subjected to several Markovian noises. They analyzed the DJRSP scheme by solving the master equation in Lindblad form.

In real-world implementation,  quantum communication protocols usually encountered four types of noise, namely the bit-flip, phase-flip (phase-damping), depolarizing, and amplitude-damping noise. In this paper, we will study noise influence of all types of noise on DJRSP. Taking a one-qubit GHZ based DJRSP scheme as an example, we will show that for different types of noise, the prepared state and the fidelity of the output state are quite different from each other. The rest of this paper is organized as follows. In Sect. 2, we show our DJRSP scheme of an arbitrary one-qubit state in ideal environment. Then, we investigate the effect of noise on the scheme with the four types of noise in Sect. 3, respectively.  The paper is concluded in Sect. 4.

\section{DJRSP of an arbitrary one-qubit state in ideal environment}
In the following, we will show a DJRSP scheme of an arbitrary one-qubit state based on GHZ state.  As we discussed in Ref. \cite{Wang-471}, this scheme is equivalent to the Bell state based scheme in Ref. \cite{NguyenCao-385}.

\subsection{DJRSP scheme of one-qubit based on GHZ state}
In our DJRSP scheme, two preparers Alice and Bob want to jointly prepare a qubit state for remote receiver Charlie. The prepared state has the form
\begin{equation} \label{EQ2}
{\left| \varphi \right\rangle} = a_0 e^{\text{i}\theta_0} {\left| 0 \right\rangle}+ a_1 e^{\text{i} \theta_1} {\left| 1\right\rangle},
\end{equation}
where $a_0, a_1\in \mathcal{R}$ with $\sum^1_{j=0} a_j^2 = 1$; $\theta_0, \theta_1 \in [0, 2\pi]$.
The information of the prepared state is split in the following way: Alice knows $S_1 = \{a_0, a_1\}$ and Bob knows $S_2=\{ \theta_0, \theta_1 \}$.
A three-qubit GHZ state is shared among Alice, Bob and Charlie as quantum resource, which has the form
\begin{equation} \label{EQ3}
{\left| \text{GHZ}_3 \right\rangle} = \frac{1}{\sqrt{2}}({\left| 000 \right\rangle}+{\left|111 \right\rangle})_{\text{ABC}},
\end{equation}
where the subscripts denote the qubits of the GHZ state. Here, Alice holds qubit A, Bob holds qubit B and Charlie holds qubit C.

Our DJRSP scheme can be described as follows.

\textbf{Step 1:} Alice performs a projective measurement on qubit A in the basis defined by $S_1$ as $\{ {\left| P_{m} \right\rangle}; m \in \{0,1\} \}$ with ${\left| P_{0}  \right\rangle} = a_{0} {\left| 0 \right\rangle} + a_{1}{\left| 1 \right\rangle}$, ${\left| P_{1}  \right\rangle} = a_{1} {\left| 0 \right\rangle} - a_{0}{\left| 1 \right\rangle}$.
Then, the quantum resource shared among three participants becomes
\begin{equation}\label{EQ4}
{\left| \text{GHZ}_3 \right\rangle}_{\text{ABC}}
 = \frac{1}{\sqrt{2}}  \sum^{1}_{m=0} {\left| P_{m} \right\rangle}_{\text{A}} {\left| Q_{m} \right\rangle}_{\text{BC}},
\end{equation}
where
${\left| Q_{0} \right\rangle}_{\text{BC}} =  a_0 {\left| 00 \right\rangle} + a_1 {\left| 11 \right\rangle}$,
${\left| Q_{1} \right\rangle}_{\text{BC}} =  a_1 {\left| 00 \right\rangle} - a_0 {\left| 11 \right\rangle}$.
After the measurement, Alice broadcasts her measurement outcome $m$ to Bob and Charlie via classical channels.

\textbf{Step 2:} 
Bob measures qubit B in the basis $\{ {\left|\right. O_n^{(m)} \left.\right\rangle}; m, n \in \{0,1\}\}$ that determined by both $S_2$ and $m$, which have the form
\begin{equation} \label{EQ5}
    \left(\begin{array}{l}
    {{\left|\right. O_{0}^{(m)} \left.\right\rangle} } \\
    {{\left|\right. O_{1}^{(m)} \left.\right\rangle} }  \end{array}\right)
= V^{(m)}
 \left(\begin{array}{l}
 {{\left| 0 \right\rangle} } \\
 {{\left| 1 \right\rangle} } \end{array}\right),
\end{equation}
with
\begin{equation}
\begin{split}
V^{(0)}=\frac{1}{\sqrt{2}}
\left(\begin{array}{cc}
 { e^{- \text{i} \theta_0} } &  { e^{- \text{i} \theta_1} } \\
 { e^{- \text{i} \theta_0} } & {- e^{- \text{i} \theta_1} }
 \end{array}\right), ~~
V^{(1)}=\frac{1}{\sqrt{2}}
\left(\begin{array}{cc}
 { e^{- \text{i} \theta_1} }  & { e^{- \text{i} \theta_0} } \\
 { -e^{- \text{i} \theta_1}} & { e^{- \text{i} \theta_0}}
 \end{array}\right).
 \end{split}
\end{equation}

After Bob performed his measurement, ${\left| Q_{m} \right\rangle}$ can be rewritten as
\begin{equation}\label{EQ6}
{\left| Q_{m} \right\rangle}_{\text{BC}}
 = \frac{1}{\sqrt{2}}  \sum^{1}_{n=0} {\left|\right. O_n^{(m)} \left.\right\rangle}_{\text{B}} {R_n^{(m)}}^{\dagger} {\left| \varphi \right\rangle}_{\text{C}},
\end{equation}
where $R_{n}^{(m)}$ denotes the recovery operator that the receiver Charlie needs to perform, which has the form $R_{0}^{(0)} = I $,  $R_{1}^{(0)} = \sigma_z$, $R_{0}^{(1)} = -\sigma_z\sigma_x$ and $R_{1}^{(1)} = -\sigma_x$.

\textbf{Step 3:} Bob announces his measurement result $n$ publicly, then Charlie can perform the recovery operator $R_{n}^{(m)}$ on qubit C to get the prepared state ${\left| \varphi \right\rangle}$.

\subsection{Density operators representation}
In quantum noisy environment, a pure state will be transformed into a mixed state, which is more convenient to be represented by density operator rather than vector state. To analyze the noisy procedure, we need to rewrite the scheme in the form of density operator. The prepared state can be written as
      \begin{equation} \label{EQ10}
       \rho_\mathrm{target} =  {\left| \varphi \right\rangle}{\left\langle \varphi \right|}.
        \end{equation}
While the quantum resource shared among three participants is
         \begin{equation} \label{EQ11}
       \rho_\mathrm{source} =   \rho_\mathrm{pure}  = {\left| \text{GHZ}_3 \right\rangle}{\left\langle \text{GHZ}_3 \right|}.
        \end{equation}
Alice's measurement operator is represented by $MA = \{MA_0, MA_1\}$, which has the form
    \begin{equation} \label{EQ12}
     MA_m =  {\left| P_m \right\rangle}{\left\langle P_m  \right|}, ~~ m \in \{0,1\}.
        \end{equation}
And Bob's measurement operator is $MB^{(m)} = \{MB_0^{(m)},MB_1^{(m)}\}$, where
    \begin{equation} \label{EQ13}
      MB_n^{(m)} =  {\left|\right. O_n^{(m)} \left.\right\rangle}{\left\langle\right. O_n^{(m)}  \left.\right|},~~n \in \{0,1\}.
        \end{equation}

Then, our DJRSP can be represented as follows.

\textbf{Step 1:}  Alice firstly measures qubit A by using the measurement operators $\{MA_m\}$ with $m \in \{0,1\}$, and the system of (B, C) will become
        \begin{equation}\label{EQ14}
        \begin{split}
        \rho_{Q_m} & = \text{tr}_{\text{A}} \left[ \frac{MA_m * \rho_\mathrm{source} * MA_m^{\dag}}{\text{tr}(MA_m^{\dag}*  MA_m * \rho_\mathrm{source})} \right].
        \end{split}
        \end{equation}

\textbf{Step 2:} Bob measures qubit B by using $\{MB_n^{(m)}\}$ with $n \in \{0,1\}$, and qubit C becomes
        \begin{equation}\label{EQ15}
        \begin{split}
        \rho_{O_n^{(m)}} & =\text{tr}_{\text{B}} \left[
        \frac{ {MB_n^{(m)}} * \rho_{Q_m}  * {MB_n^{(m)}}^{\dag}}
           { \text{tr} \left( {MB_n^{(m)}}^{\dag}* {MB_n^{(m)}} *  \rho_{Q_m} \right)} \right]
        .
        \end{split}
        \end{equation}

\textbf{Step 3:} Charlie recover the prepared state by performing $R_n^{(m)}$, that is
       \begin{equation}\label{EQ16}
        \begin{split}
        \rho_\mathrm{out} & = R_n^{(m)} * \rho_{O_n^{(m)}} * {R_n^{(m)}}^{\dag} = \rho_\mathrm{target}.
        \end{split}
        \end{equation}

\section{DJRSP of an arbitrary one-qubit state in noisy environment}
In ideal situation, it is assumed that an entangled quantum resource has been shared among three participants. However, in real situation, there must be a source that generates the entangled states and distributes each qubit to relevant participant. And each distribution quantum channel will inevitably be affected by quantum noise in real-world implementation. In the following, we will discuss how the noise around distribution channels affects the DJRSP scheme.

\subsection{The noise channels}
There are four types of noise usually encountered in real-world quantum communication protocols, namely the bit-flip, phase-flip (phase-damping), depolarizing and amplitude-damping noise.

\subsubsection{The bit-flip noise}
The bit-flip noise changes the state of a qubit from ${\left| 0 \right\rangle}$ to ${\left| 1 \right\rangle}$ or from ${\left| 1 \right\rangle}$ to ${\left| 0 \right\rangle}$ with probability $\lambda$ and its Kraus operators are \cite{Xian-Ting-940}
\begin{equation} \label{EQ31}
        E_0= \sqrt{1-\lambda}~I,
        ~~
        E_1= \sqrt{\lambda}~\sigma_x,
\end{equation}
where $I$ is identity matrix,  $\sigma_x$ is the Pauli matrix and $0 \leq \lambda \leq 1$ is the noise parameter.

\subsubsection{The phase-flip (phase-damping) noise}
The phase-flip noise changes the phase of the qubit ${\left| 1 \right\rangle}$  to $-{\left| 1 \right\rangle}$ with probability $\lambda$ and it can be described by Kraus operators as \cite{Xian-Ting-940}
\begin{equation} \label{EQ32}
        E_0= \sqrt{1-\lambda}~I,
        ~~
        E_1= \sqrt{\lambda}~\sigma_z,
\end{equation}
where $\sigma_z$ is the Pauli matrix and $0 \leq \lambda \leq 1$. Note that the phase-flip noise is equivalent to the phase-damping noise, which describes the loss of quantum information without energy dissipation.

\subsubsection{The depolarizing noise}
The depolarizing noise takes a qubit and replaces it with a completely mixed state $I/2$ with probability $\lambda$ and its Kraus operators are \cite{Xian-Ting-940}
\begin{equation} \label{EQ33}
\begin{split}
        E_0= \sqrt{1-\lambda}~I, ~~
        E_1= \sqrt{\frac{\lambda}{3}}~\sigma_x, ~~
        E_2= \sqrt{\frac{\lambda}{3}}~\sigma_z, ~~
        E_3= \sqrt{\frac{\lambda}{3}}~\sigma_y,
\end{split}
\end{equation}
where $\sigma_x, \sigma_z, \sigma_y$ are Pauli matrices and $0 \leq \lambda \leq 1$.

\subsubsection{The amplitude-damping noise}
The amplitude-damping noise describes the energy dissipation effects due to loss of energy from a quantum system and its Kraus operators are as follows \cite{Xian-Ting-940}
\begin{equation} \label{EQ34}
        E_0=
        \left( \begin{array}{cc}
            1 & 0 \\
            0 & \sqrt{1-\lambda}
          \end{array}\right),
          ~~
          E_1=
        \left(\begin{array}{cc}
          0   &  \sqrt{\lambda} \\
          0  & 0
         \end{array}\right),
\end{equation}
where $0 \leq \lambda \leq 1$ indicates the noise parameter.

\subsection{The output state and the fidelity in noise environment}
\label{sec:3-2}
Suppose Alice has a quantum source generator in her laboratory. She generates the entangled resource ${\left| \text{GHZ}_3 \right\rangle}_{\text{ABC}}$, keeps qubit A in her own and then sends B to Bob and C to Charlie via noisy quantum channels, respectively. To simplify the analysis, we suppose that the noise type of each channel is identical. In this case, the entangled source shared among three participants after qubits transmission can be rewritten as
  \begin{equation}
  \begin{split}
  \rho_\mathrm{source}
         & =\epsilon( \rho_\mathrm{pure} ) \\
         & = \sum_{j_1, j_2}
         { E_{j_1}^{(\text{B})}  E_{j_2}^{(\text{C})} }  \left| \text{GHZ}_3 \right\rangle  \left\langle \text{GHZ}_3 \right|
         {E_{j_1}^{(\text{B})}}^{\dag}  {E_{j_2}^{(\text{C})}}^{\dag},
        \end{split}
         \end{equation}
where $E_{j_1}, E_{j_2}$ represent the noise operators that act on different qubits and superscripts denote the qubit transmitted through noise channel.
And the fidelity of the output state can be calculated as
\begin{equation}
\begin{split}
F := | {\left\langle \varphi \right|} \rho_\mathrm{out}  {\left| \varphi \right\rangle}|.
\end{split}
\end{equation}

To analyze noise effect of each type of noise, we just need to recalculate $\rho_{source}$, put it into Eq. (\ref{EQ14}), and get results from Eqs. (\ref{EQ15}) and (\ref{EQ16}). For the above four types of noise, we will get the following results.

\subsubsection{DJRSP of one-qubit in the bit-flip noise}
In the bit-flip noise, we will get
\begin{equation}
\begin{split}
      \rho_\mathrm{out}^\mathrm{BF} = &[a_0^2 (1-2 \lambda )+\lambda]\left| 0\right\rangle \left\langle 0\right|
     +[a_1^2 (1-2 \lambda )+\lambda]\left| 1\right\rangle \left\langle 1\right|  \\
  &+ a_0 a_1\left| 0\right\rangle \left\langle 1\right|
\left[ e^{\text{i} ( \theta _0 - \theta _1)}(2 \lambda ^2-2 \lambda +1) - 2 e^{\text{i} (\theta _1 -\theta _0)}\lambda(\lambda -1) \right]
\\&+ a_0 a_1\left| 1\right\rangle \left\langle 0\right|
\left[ e^{\text{i} (\theta _1-\theta _0)} (2 \lambda ^2-2 \lambda +1)- 2 e^{\text{i} \left(\theta _0-\theta _1\right)}\lambda (\lambda -1) \right] .
\end{split}
\end{equation}
And the fidelity is
\begin{equation}
\begin{split}
      F^\mathrm{BF} = 1-\lambda  + 4 \lambda  (a_1^2 - a_1^4)\left[ \lambda +(1-\lambda) \cos (2 \theta _0-2\theta _1) \right].
  \end{split}
  \end{equation}

\begin{figure}
\centering
\subfigure[]{
\begin{minipage}[t]{0.3\textwidth}
\centering
\includegraphics [scale=0.25]{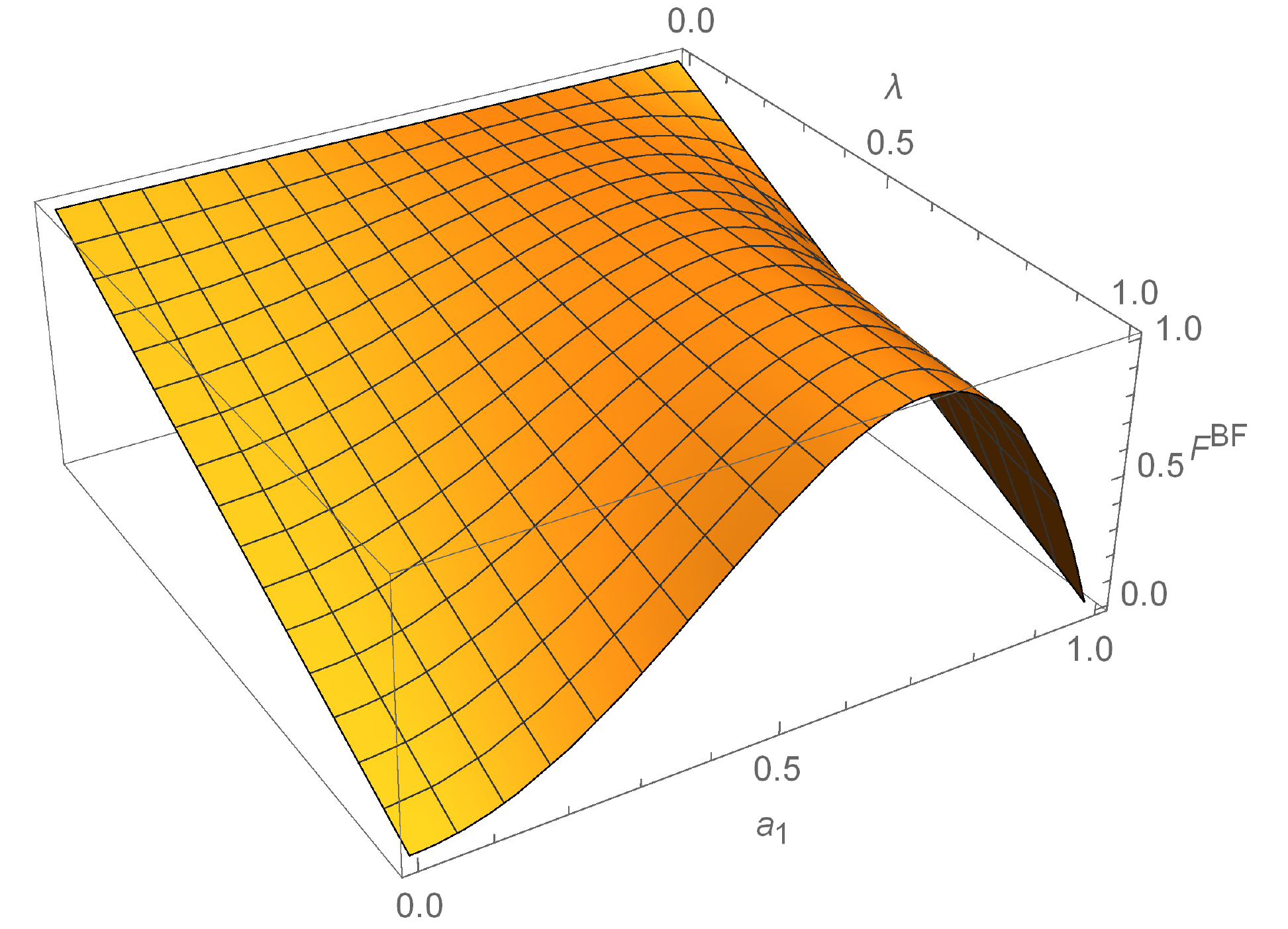}\label{fig:2:a1}
\end{minipage}}
\hspace{0.02\textwidth}
\subfigure[]{
\begin{minipage}[t]{0.3\textwidth}
\centering
\includegraphics [scale=0.25]{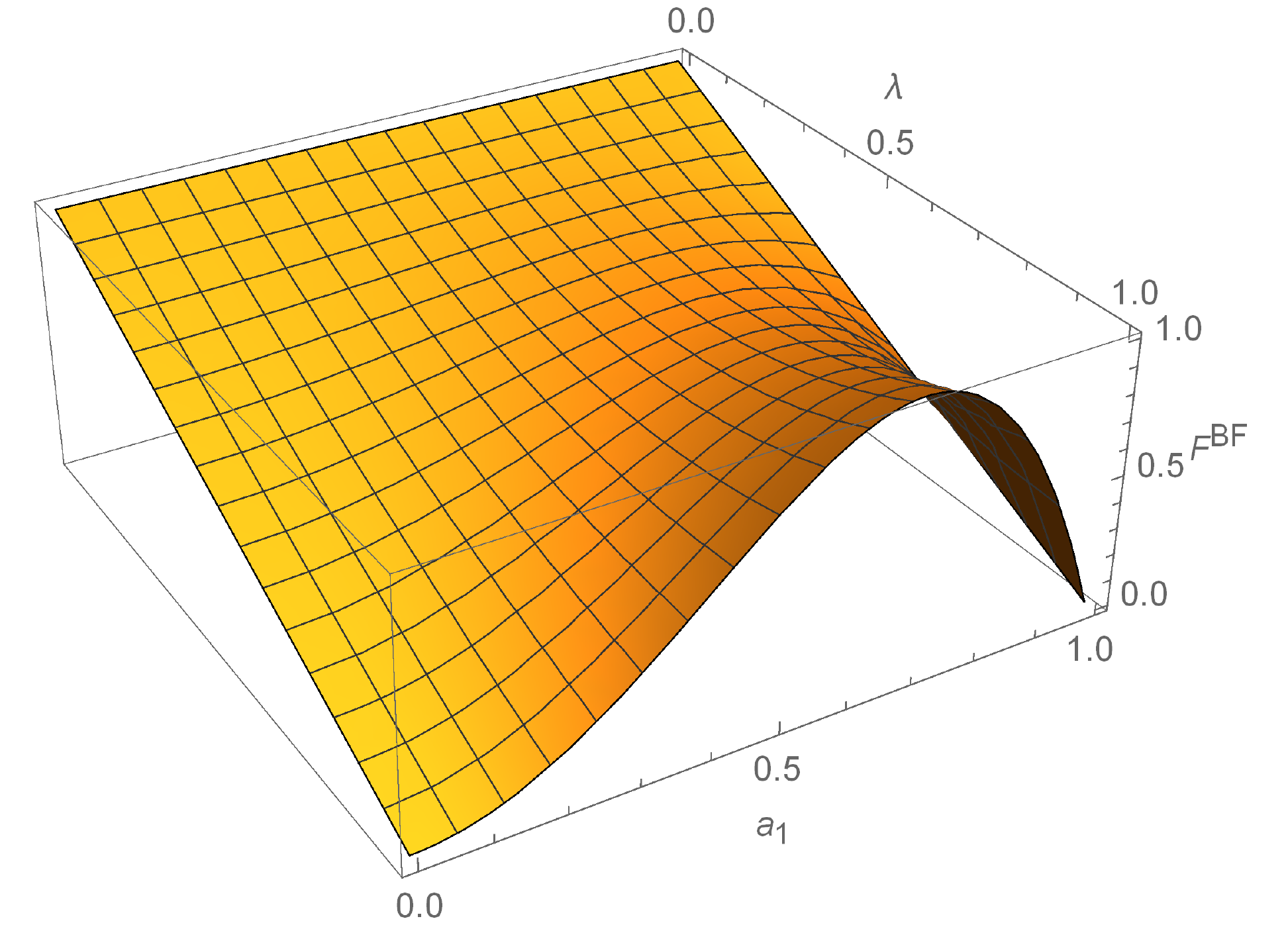}\label{fig:2:a2}
\end{minipage}}
\hspace{0.02\textwidth}
\subfigure[]{
\begin{minipage}[t]{0.3\textwidth}
\centering
\includegraphics [scale=0.25]{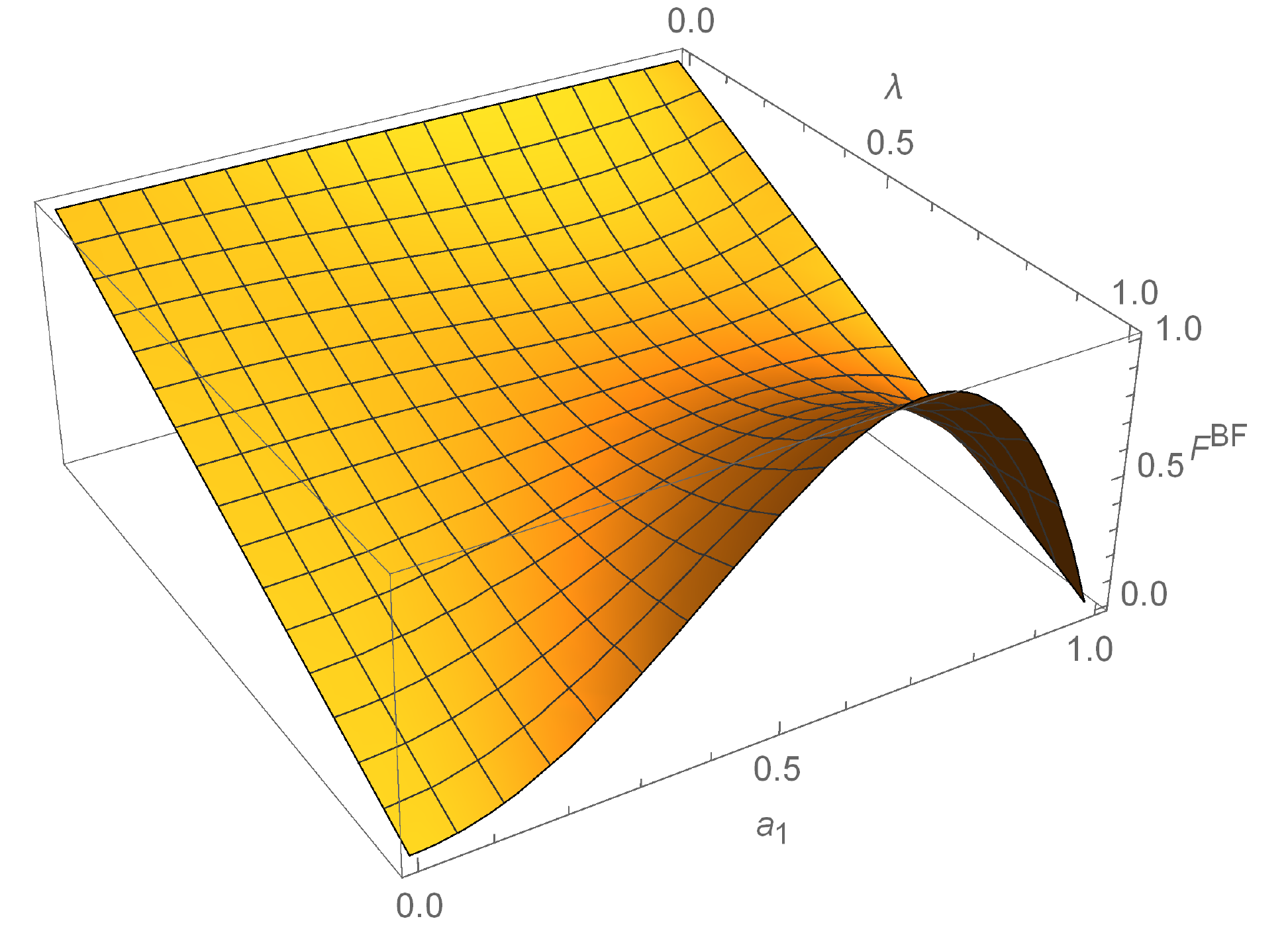}\label{fig:2:a3}
\end{minipage}}
\hspace{0.02\textwidth}
\subfigure[]{
\begin{minipage}[t]{0.3\textwidth}
\centering
\includegraphics [scale=0.22]{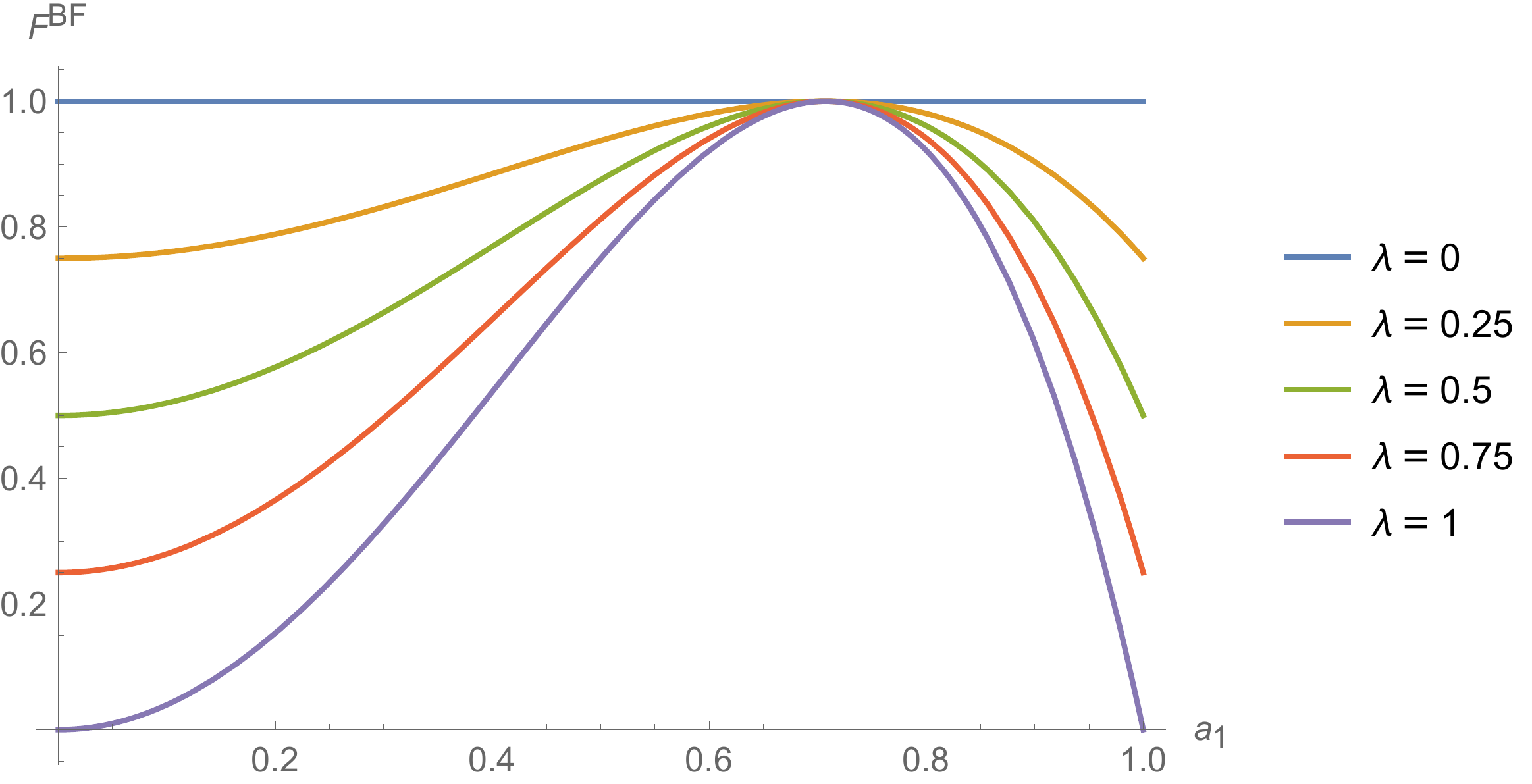}\label{fig:2:b1}
\end{minipage}}
\hspace{0.02\textwidth}
\subfigure[]{
\begin{minipage}[t]{0.3\textwidth}
\centering
\includegraphics [scale=0.22]{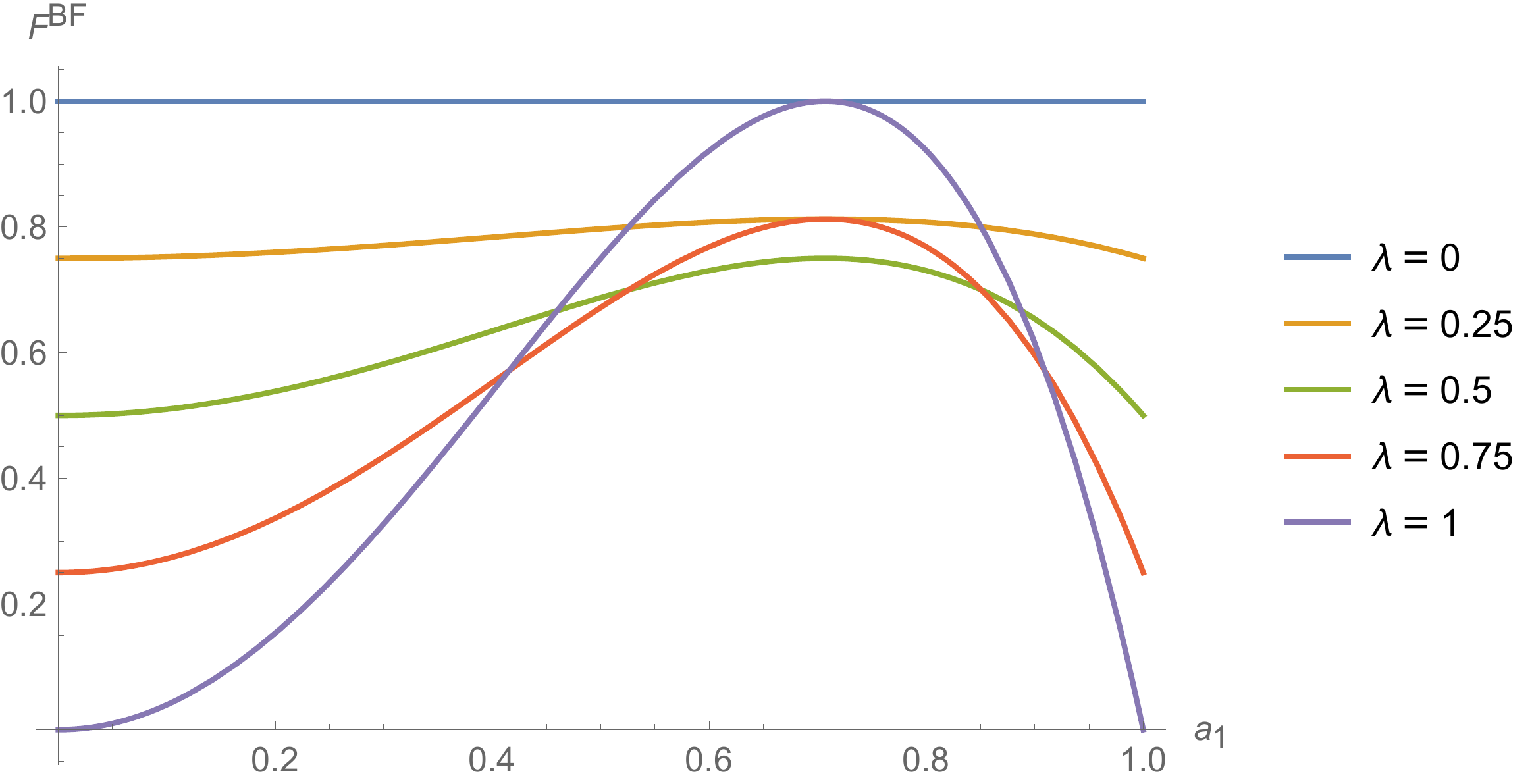}\label{fig:2:b2}
\end{minipage}}
\hspace{0.02\textwidth}
\subfigure[]{
\begin{minipage}[t]{0.3\textwidth}
\centering
\includegraphics [scale=0.22]{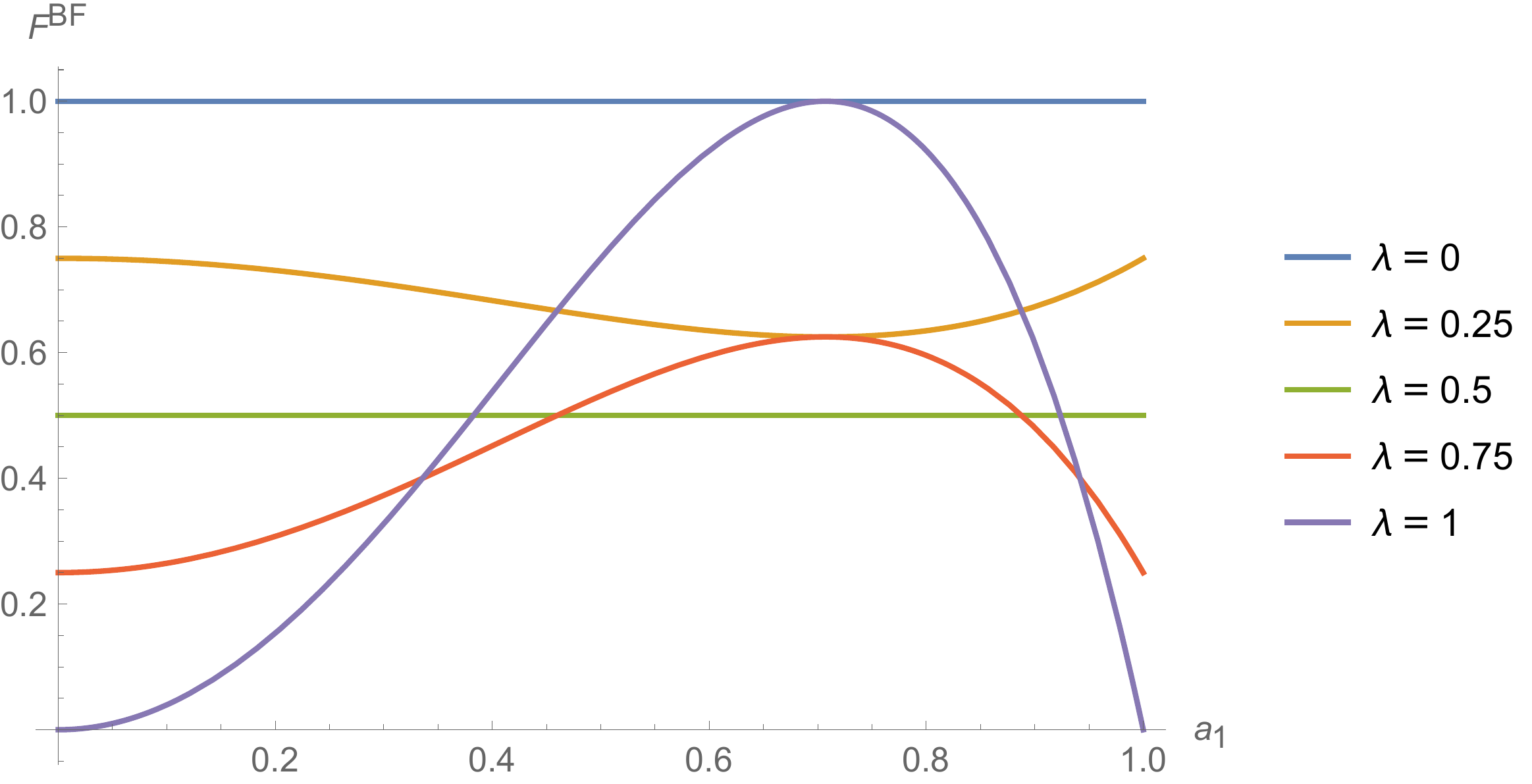}\label{fig:2:b3}
\end{minipage}}
\caption{The fidelity $F^\mathrm{BF}$ of the output state in the bit-flip noise with respect to $\lambda$ and $a_1$ for different values of $\theta _0 - \theta _1$. (a) $F^\mathrm{BF}$  with $\lambda$ and $a_1$ when $\theta _0 - \theta _1 = 0$ or $\pi$ ; (b) $\theta _0 - \theta _1 = \frac{\pi}{4}$ or $ \frac{3\pi}{4}$; (c) $\theta _0 - \theta _1 = \frac{\pi}{2}$; (d) $F^\mathrm{BF}$ with  $a_1$ for selected $\lambda$ when $\theta _0 - \theta _1 = 0$ or $\pi$ ;  (e) $F^\mathrm{BF}$ with  $a_1$ for selected $\lambda$ when $\theta _0 - \theta _1 = \frac{\pi}{4}$ or $ \frac{3\pi}{4}$; (f) $F^\mathrm{BF}$ with  $a_1$ for selected $\lambda$ when $\theta _0 - \theta _1 = \frac{\pi}{2}$.}
\label{fig:2}
\end{figure}

It can be seen from the above equation that the fidelity is relevant to the noise rate $\lambda$, the amplitude factor $a_1$ ($a_0 = \sqrt{1- a_1^2}$), and also the phase factors $\theta _0$ and $\theta _1$. The relationship of $F^\mathrm{BF}$, $\lambda$ and $a_1$ with different values of $\theta _0 - \theta _1$ are plotted in Fig. \ref{fig:2}.

Fig. \ref{fig:2:a1} represents  $F^\mathrm{BF}$  with $\lambda$ and $a_1$  in the case of $\theta _0 - \theta _1 = 0$ or $\pi$. As shown, the maximum fidelity is 1 when $\lambda =0$ or $a_1=\frac{1}{\sqrt{2}}$, which means there is no noise or the prepared state is $ \frac{{\left| 0 \right\rangle}\pm {\left| 1 \right\rangle}}{\sqrt{2}}$ that is immune to the bit-flip noise.
The minimum fidelity is 0 when $\lambda =1$ and $a_1 = 0 ~\text{or} ~1$, which means the prepared state is ${\left| 0 \right\rangle}$ or ${\left| 1 \right\rangle}$ and the bit-flip noise will change the prepared state to its orthogonal state, i.e., ${\left| 0 \right\rangle} \rightarrow {\left| 1 \right\rangle}$ or ${\left| 1 \right\rangle} \rightarrow {\left| 0 \right\rangle}$.
If $\lambda$ takes some certain values, we can get related curves in \ref{fig:2:b1}, which are specific instances of the surface in \ref{fig:2:a1}. It can be seen from the figure that the fidelity is convex upward and it changes dramatically with the increase of noise rate $\lambda$. $F^\mathrm{BF}$ will get the maximum point 1 if  $a_1 = \frac{1}{\sqrt{2}}$ for all   $\lambda$.

For other cases where $\theta _0 - \theta _1  = \frac{\pi}{4}$, $ \frac{3\pi}{4}$ or $ \frac{\pi}{2}$, the surfaces are plotted in Figs. \ref{fig:2:a2} and \ref{fig:2:a3}. While \ref{fig:2:b2} and \ref{fig:2:b3} are specific examples of \ref{fig:2:a2} and \ref{fig:2:a3} when $\lambda$  takes some certain values, respectively. 
It can be seen from \ref{fig:2:b2} and \ref{fig:2:b3} that in the case of $0<\lambda<1$, $F^\mathrm{BF}$ is always less than 1 no matter what value $a_1$ is.

\subsubsection{DJRSP of one-qubit in the phase-flip noise}
\begin{figure}
\centering
\subfigure[]{
\begin{minipage}[t]{0.3\textwidth}
\centering
\includegraphics [scale=0.27]{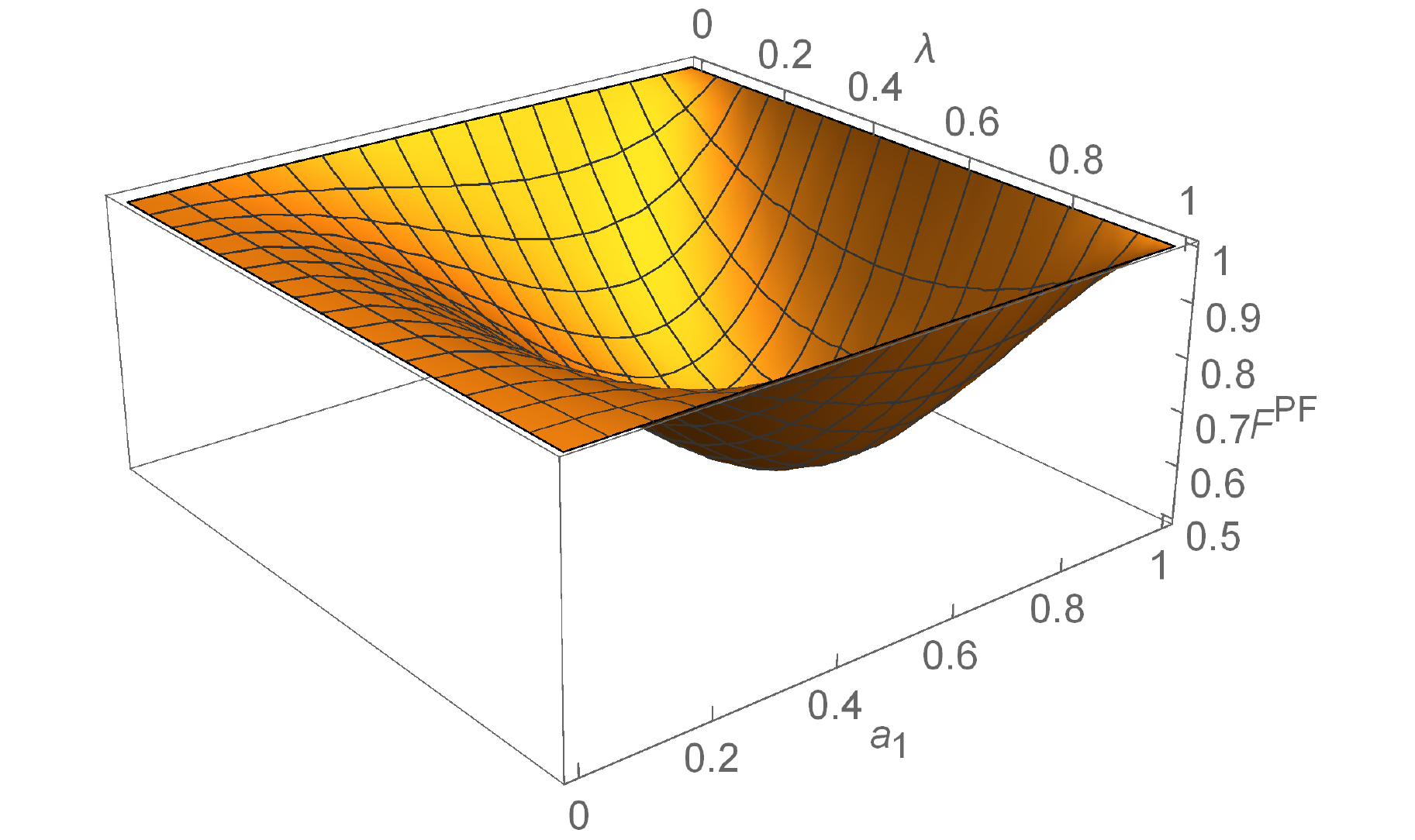}\label{fig:3:a}
\end{minipage}}
\hspace{0.02\textwidth}
\subfigure[]{
\begin{minipage}[t]{0.3\textwidth}
\centering
\includegraphics[scale=0.22]{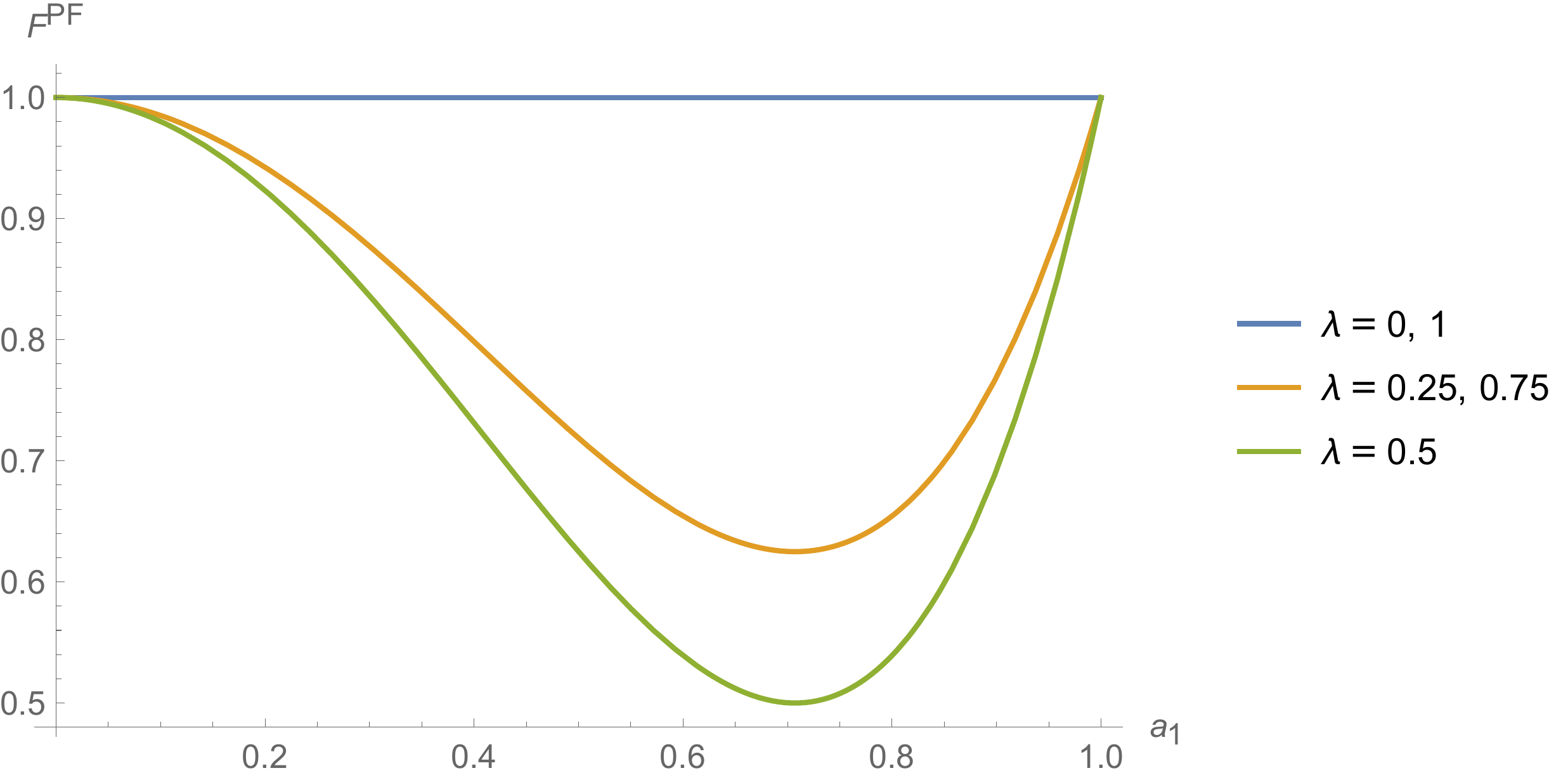}\label{fig:3:b}
\end{minipage}}
\caption{The fidelity $F^\mathrm{PF}$ of the output state in the phase-flip noise with respect to $\lambda$ and $a_1$. (a) $F^\mathrm{PF}$ with $\lambda$ and $a_1$; (b) $F^\mathrm{PF}$ with $a_1$ for some selected values of $\lambda$.}
\label{fig:3}
\end{figure}

In the phase-flip noise, we have
  \begin{equation}
  \begin{split}
  \rho_\mathrm{out}^\mathrm{PF}  = &
   a_0^2 \left| 0\right\rangle  \left\langle 0\right|
+a_1^2 \left| 1\right\rangle  \left\langle 1\right|
+a_0 a_1(1-2 \lambda )^2 \left[
e^{\text{i} \left(\theta _0-\theta _1\right)} \left| 0\right\rangle  \left\langle 1\right|
+e^{\text{i} \left(\theta _1-\theta _0\right)}\left| 1\right\rangle  \left\langle 0\right|   \right].
        \end{split}
         \end{equation}
And the fidelity is
\begin{equation}
\begin{split}
F ^\mathrm{PF}   =  1 - 8  \lambda (1-\lambda) (a_1^2- a_1^4).
\end{split}
\end{equation}
Note that the fidelity is relevant to the noise rate $\lambda$ and the amplitude factor $a_1$, but not the phase factors $\theta _0$ and $\theta _1$, which is different from the bit-flip noise. The relationship of $F^\mathrm{PF}$, $\lambda$ and $a_0$ is shown in Fig. \ref{fig:3}.
As shown, the maximum fidelity is 1 when $\lambda =0$, or $\lambda =1$, or $a_1=0$ or $a_1=1$, which means there is no noise, or the noise does not change the entanglement, or the prepared state is ${\left| 0 \right\rangle}$ or $ {\left| 1 \right\rangle}$.
The minimum fidelity is $\frac{1}{2}$ when $\lambda = \frac{1}{2}$ and $a_1 = \frac{1}{\sqrt{2}}$, which means the prepared state is $ \frac{e^{\text{i}\theta_0} {\left| 0 \right\rangle}+ e^{\text{i} \theta_1} {\left| 1\right\rangle}}{\sqrt{2}}$ and the output state is complete mixture $ \frac{ \left| 0\right\rangle  \left\langle 0\right|+ \left| 1\right\rangle \left\langle 1\right| }{2}$.

The value of  $F^\mathrm{PF}$ with $a_1$  is presented in Fig. \ref{fig:3:b} for different $\lambda$.  It can be seen from the figure that the fidelity is concave upward if $\lambda \neq 1$. The fidelities will be the same if $\lambda$ is set to $x$ and $1-x$ with $0 \leq x\leq1$ (for example, the fidelities are the same when $\lambda = 0.25$ and $\lambda = 0.75$). And each $F^\mathrm{PF}$ will get its minimum point if  $a_1 = \frac{1}{\sqrt{2}}$ for all  $\lambda$.

\subsubsection{DJRSP of one-qubit in the depolarizing noise}
\begin{figure}
\centering
\subfigure[]{
\begin{minipage}[t]{0.3\textwidth}
\centering
\includegraphics[scale=0.22]{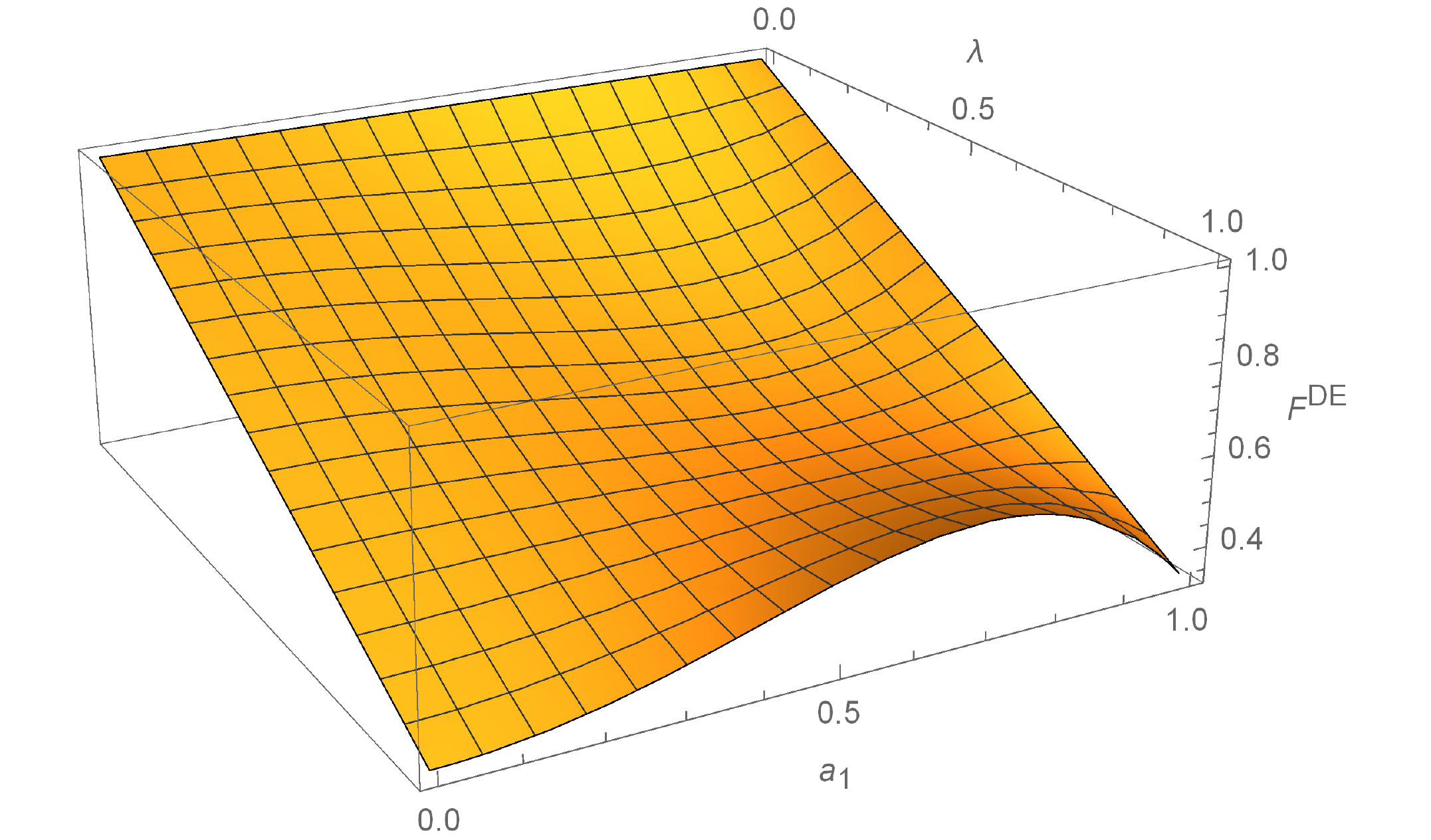}\label{fig:4:a}
\end{minipage}}
\hspace{0.02\textwidth}
\subfigure[]{
\begin{minipage}[t]{0.3\textwidth}
\centering
\includegraphics[scale=0.25]{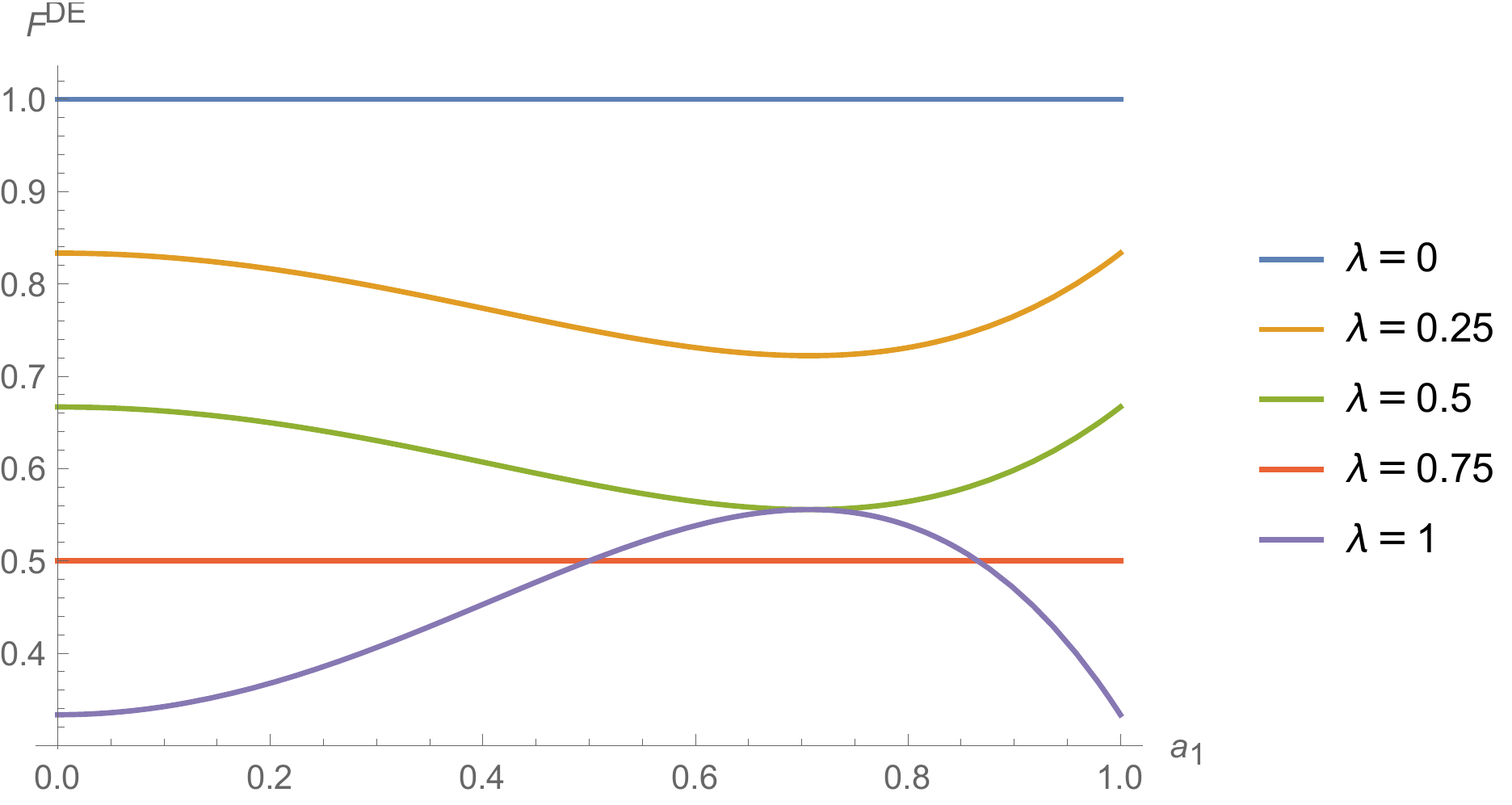}\label{fig:4:b}
\end{minipage}}
\caption{
The fidelity $F^\mathrm{DE}$ of the output state in the depolarizing noise with respect to $\lambda$ and $a_1$. (a) $F^\mathrm{DE}$ with $\lambda$ and $a_1$; (b) $F^\mathrm{DE}$ with $a_1$ for some selected values of $\lambda$.}
\label{fig:4}
\end{figure}

In the depolarizing noise, the output state is
  \begin{equation}
  \begin{split}
  \rho_\mathrm{out}^\mathrm{DE} = &
   \frac{1}{3} [a_0^2 (3-4 \lambda )+2\lambda]  \left| 0\right\rangle  \left\langle 0\right|
 +\frac{1}{3}[a_1^2 (3-4 \lambda )+2 \lambda] \left|1 \right\rangle  \left\langle 1\right|  \\ &
  +\frac{1}{9}a_0 a_1 e^{\text{i} (\theta _0-\theta _1)} (3-4 \lambda )^2 \left| 0\right\rangle  \left\langle 1\right|
  +\frac{1}{9}a_0 a_1 e^{\text{i} (\theta _1-\theta _0)}(3-4 \lambda )^2 \left|1\right\rangle  \left\langle 0\right|.
  \end{split}
  \end{equation}
And the fidelity is
\begin{equation}
\begin{split}
      F^\mathrm{DE} =  1-\frac{2}{3} \lambda + \frac{8}{9}  \lambda  (4 \lambda -3) ( a_1^2 - a_1^4 ),
  \end{split}
  \end{equation}
where the fidelity is still relevant to the noise rate $\lambda$ and the amplitude factor $a_1$. The relationship of $F^\mathrm{DE}$, $\lambda$ and $a_1$ is shown in Fig. \ref{fig:4}. As shown, the maximum fidelity is 1 when $\lambda =0$, which means there is no noise.  The minimum fidelity is $\frac{1}{3}$ when $\lambda = 1$ and $a_1 = 0 ~\text{or} ~1$, which means the prepared state is  $ {\left| 0 \right\rangle}$ or ${\left| 1\right\rangle}$  and the output state is $ \frac{ \left| 0\right\rangle  \left\langle 0\right|+ 2 \left| 1\right\rangle \left\langle 1\right| }{3}$ or $ \frac{ 2 \left| 0\right\rangle  \left\langle 0\right|+  \left| 1\right\rangle \left\langle 1\right| }{3}$.

It can be seen from Fig. \ref{fig:4:b} that the fidelity is constant, $F^\mathrm{DE}=\frac{1}{2}$, when $\lambda =\frac{3}{4}$. The fidelity is concave upward  if $\lambda >\frac{3}{4}$ and the fidelity is convex upward if $\lambda <\frac{3}{4}$. For each curve, $F^\mathrm{DE}$ will get the maximum/minimum point when $a_1 = \frac{1}{\sqrt{2}}$.

\subsubsection{DJRSP of one-qubit in the amplitude-damping noise}
In the amplitude-damping noise, we will get two different output states based on Alice's measurement result $m$, which is different from the other three types of noise. For $m=0$, we will get the output state as
  \begin{equation}
  \begin{split}
  \rho_\mathrm{out}^\mathrm{AD_0} =   &
         [a_0^2 (1-\lambda )+\lambda] \left| 0\right\rangle  \left\langle 0\right|
         +a_1^2 (1-\lambda)       \left| 1 \right\rangle  \left\langle 1\right|    \\ &
         +a_0 a_1 (1-\lambda) e^{\text{i} (\theta _0-\theta _1)}  \left| 0\right\rangle  \left\langle 1\right|
         +a_0 a_1 (1-\lambda) e^{\text{i} (\theta _1-\theta _0)}  \left| 1\right\rangle  \left\langle 0\right| .
        \end{split}
         \end{equation}
And the corresponding fidelity is
\begin{equation}
\begin{split}
      F^\mathrm{AD_0} = 1-a_1^2 \lambda.
  \end{split}
  \end{equation}

While for $m=1$, we will get
  \begin{equation}
  \begin{split}
  \rho_\mathrm{out}^\mathrm{AD_1} =   &
         a_0^2 (1-\lambda )\left| 0\right\rangle  \left\langle 0\right|
         +[a_1^2 (1-\lambda)+\lambda]  \left| 1 \right\rangle  \left\langle 1\right|      \\ &
         +a_0 a_1 (1-\lambda) e^{\text{i} (\theta _0-\theta _1)}  \left| 0\right\rangle  \left\langle 1\right|
         +a_0 a_1 (1-\lambda) e^{\text{i} (\theta _1-\theta _0)}  \left| 1\right\rangle  \left\langle 0\right|.
        \end{split}
         \end{equation}
And the fidelity is
\begin{equation}
\begin{split}
      F^\mathrm{AD_1} = 1+ a_1^2 \lambda -\lambda. 
  \end{split}
  \end{equation}

\begin{figure}
\centering
\subfigure[]{
\begin{minipage}[t]{0.3\textwidth}\centering
\includegraphics[scale=0.28]{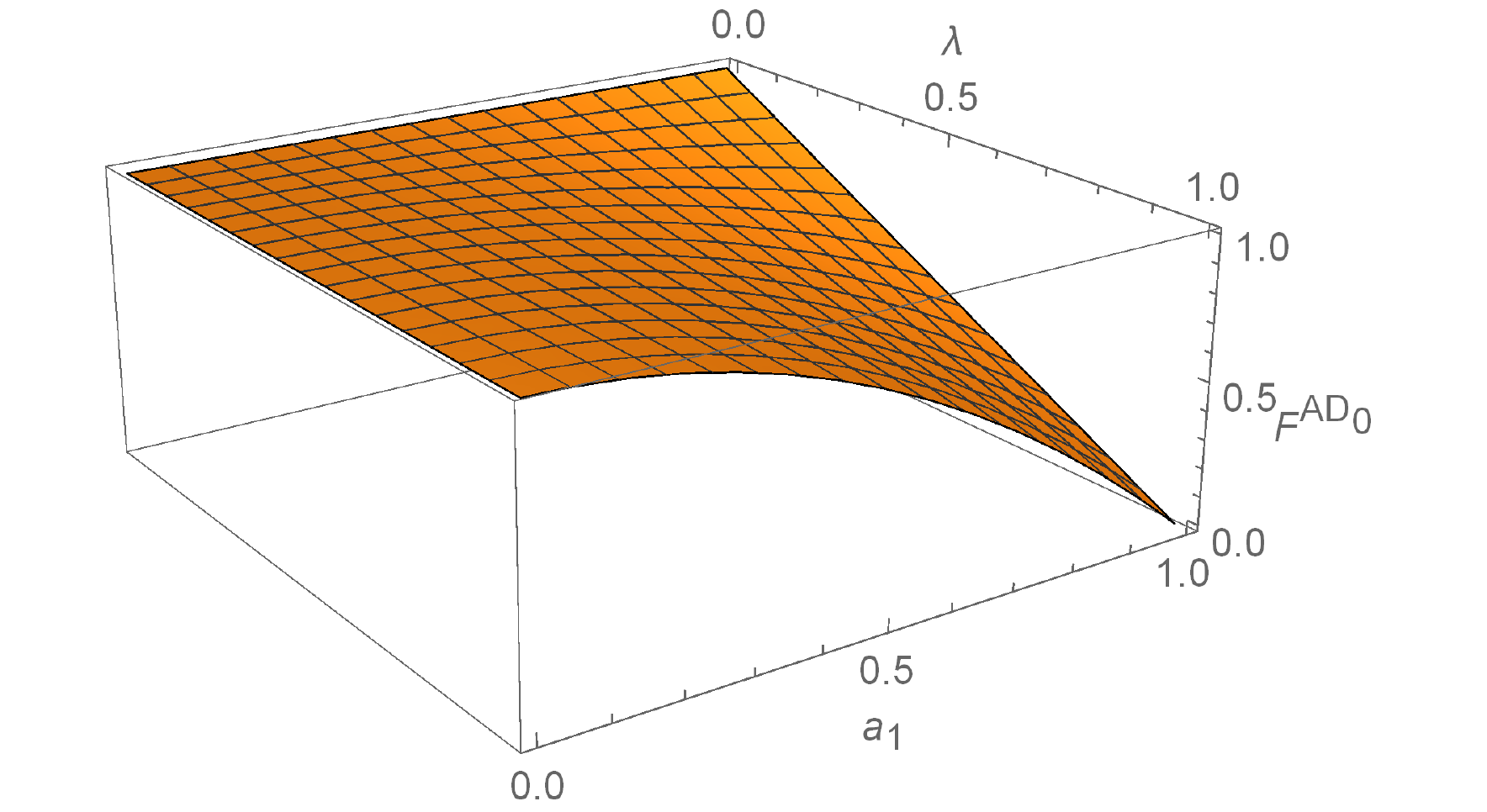}\label{fig:5:a}
\end{minipage}}
\hspace{0.05\textwidth}
\subfigure[]{
\begin{minipage}[t]{0.3\textwidth}\centering
\includegraphics[scale=0.28]{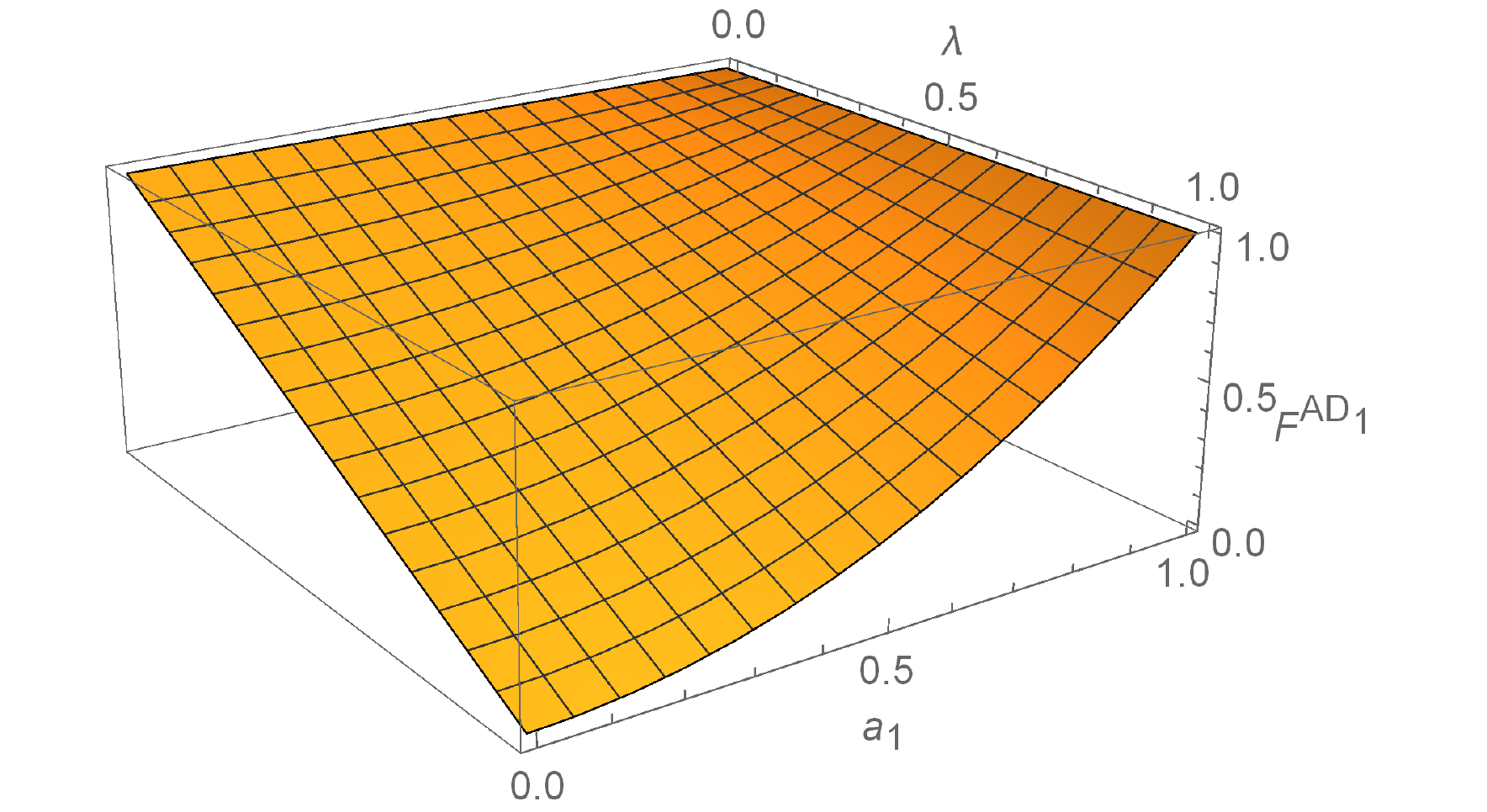}\label{fig:5:b}
\end{minipage}}
\hspace{0.05\textwidth}
\subfigure[]{
\begin{minipage}[t]{0.3\textwidth}\centering
\includegraphics[scale=0.2]{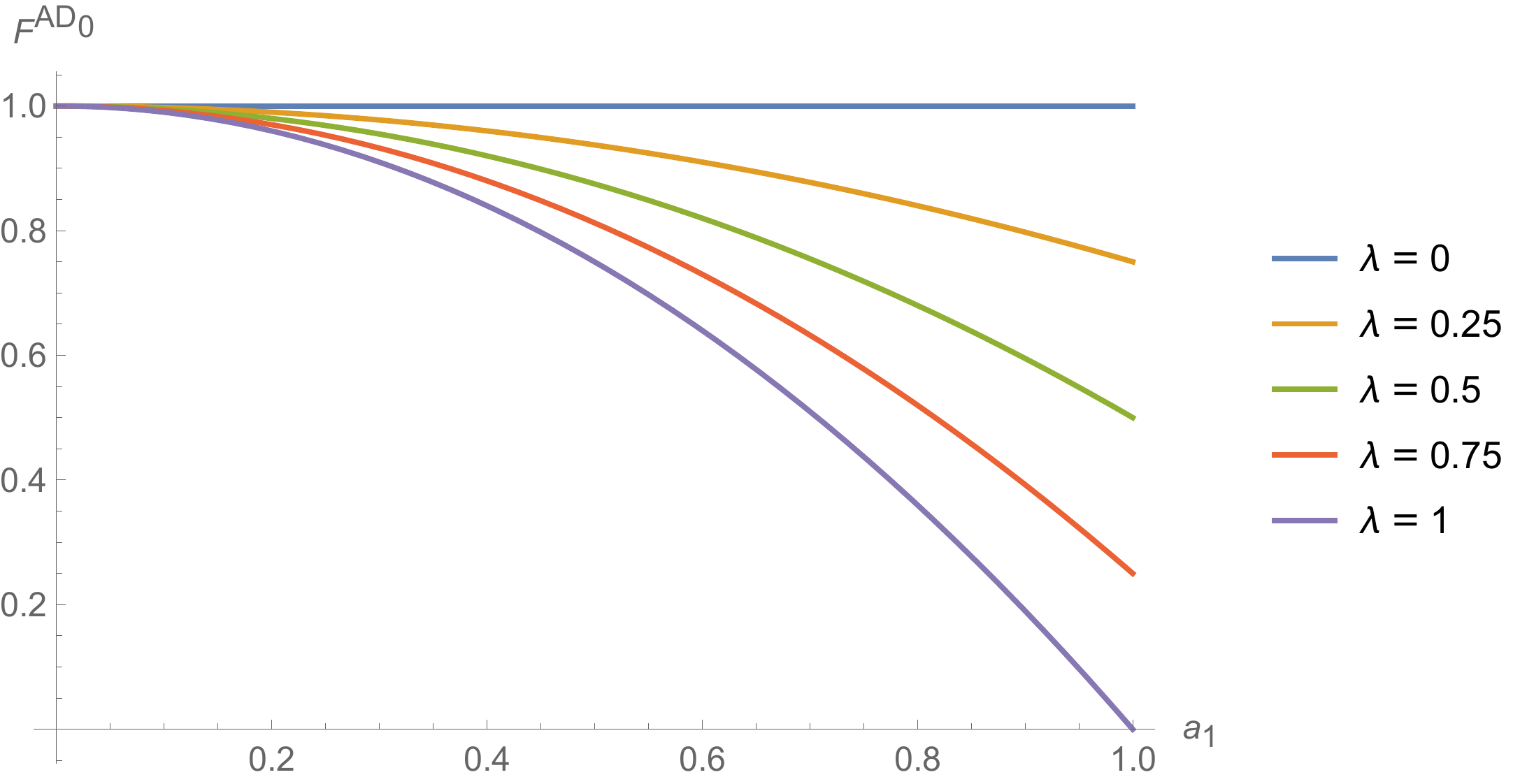}\label{fig:5:a2}
\end{minipage}}
\hspace{0.05\textwidth}
\subfigure[]{
\begin{minipage}[t]{0.3\textwidth}\centering
\includegraphics[scale=0.2]{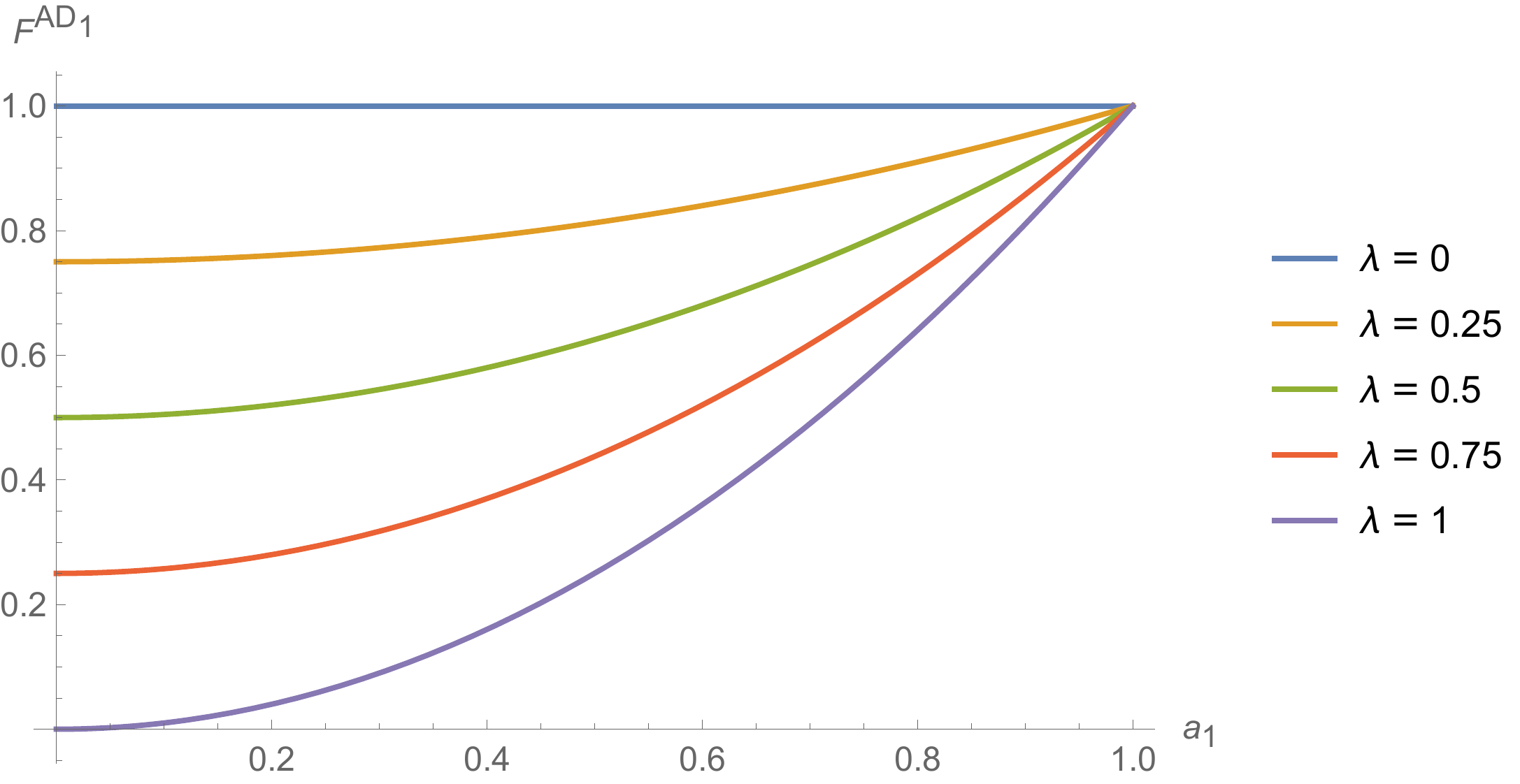}\label{fig:5:b2}
\end{minipage}}
\caption{
The fidelity of the output state in the amplitude-damping noise with respect to $\lambda$ and $a_1$.
(a) $F^\mathrm{AD_0}$ with $\lambda$ and $a_1$;
(b) $F^\mathrm{AD_1}$ with $\lambda$ and $a_1$;
(c) $F^\mathrm{AD_0}$ with $a_1$ for some selected values of $\lambda$;
(d) $F^\mathrm{AD_1}$ with $a_1$ for some selected values of $\lambda$.}
\label{fig:5}
\end{figure}

Still, the fidelity for both cases is relevant to the noise rate $\lambda$  and the amplitude factor $a_1$.
The relationship of $F^\mathrm{AD_0}$ and $F^\mathrm{AD_1}$ with $\lambda$ and $a_1$ can be found in Fig. \ref{fig:5}.
As shown in Fig. \ref{fig:5:a}, the maximum $F^\mathrm{AD_0}$ is 1 when $\lambda =0$ or $a_1=0$, which means there is no noise or the prepared state is $ {\left| 0 \right\rangle}$.
The minimum $F^\mathrm{AD_0}$  is $0$ when $\lambda =1$ and $a_1 =1$, which means the prepared state is ${\left| 1\right\rangle}$  and the output state is $\left| 0\right\rangle$.
Note that one will get the same surface of $F^\mathrm{AD_1}$ as $F^\mathrm{AD_0}$ if the variable $a_0$ is replaced by $a_1$. And similar results about the maximum and minimum $F^\mathrm{AD_1}$ can be got in Fig. \ref{fig:5:b}.

For some selected values of $\lambda$, one can get related curves in Fig. \ref{fig:5:a2} and Fig. \ref{fig:5:b2}. It can be seen from the figures that $F^\mathrm{AD_0}$ and $F^\mathrm{AD_1}$  are monotone. $F^\mathrm{AD_0}$ is convex upward and it decreases dramatically with the increase of noise rate $\lambda$ from $a_1 = 0$  to $a_1 = 1$ and each curve will get its minimum value at the right point. While $F^\mathrm{AD_1}$ is concave upward and it increases dramatically with the increase of noise rate $\lambda$ from $a_1 = 0$  to $a_1 = 1$ and each curve will get its minimum value at the left point.

\section{Conclusion}
Starting with the scheme in ideal condition, we investigated the DJRSP scheme in four types of noise, respectively. As shown in the paper, some information of the prepared state is lost through the noise channels. We use fidelity to describe how close are the final states to the original state and how much information has been lost in the process. The result of our study shows that the prepared state and the fidelity of the state is quite different from each other in different types of noise. For one thing, the fidelity of the prepared state in the bit-flip noise depends on the amplitude factor $a_i$ and the phase factor $\theta_i$ of the initial state, and the noise parameter $\lambda$. But in the other three types of noise, the fidelity only depends on the amplitude factor and the noise parameter, but have nothing to do with the phase parameter $\theta_i$. For another thing, in the amplitude-damping noise, it is interesting that the receiver Charlie will get different prepared output states depending on the first preparer Alice's measurement result $m$. But in the other three types of noise, the receiver will get the same output state, which is irrelevant to the first preparer Alice's measurement result.

We have considered the case where the qubits in Bob's and Charlie's side were affected by quantum noise. It should be noted that the qubit A in Alice's side may still be affected by noise. In this case, the noise effect on the quantum channel can be represented as
  \begin{equation}
  \begin{split}
  \rho_\mathrm{source}
          = &\epsilon( \rho_\mathrm{pure} ) \\
          = &\sum_{j_1, j_2, j_3}
         E_{j_1}^{(\text{A})} E_{j_2}^{(\text{B})}  E_{j_3}^{(\text{C})}  ~\rho_{pure}~  {E_{j_1}^{(\text{A})}}^{\dag}  {E_{j_2}^{(\text{B})}}^{\dag} {E_{j_3}^{(\text{C})}}^{\dag}.
  \end{split}
\end{equation}
We can still calculate the noise effect on entanglement channel in different types of noise, just as mentioned Sect. \ref{sec:3-2}. And it is also possible to consider the situation where different qubits are subjected to different types of noise.

In summary, we have studied a DJRSP scheme of an arbitrary single qubit in noisy environment and shown how the scheme is affected by all types of noise usually encountered in real-world. Our results will be helpful for analyzing and improving quantum secure communication in real implementation. To show our method, we have considered a simple case where three participants were involved. In the future, it is also possible to analyze other situations such as multi-participants involved or multi-qubit prepared.

\section*{Acknowledgements}
We thank the anonymous reviewers for their helpful comments. This project was supported by NSFC (Grant Nos. 61601358, 61373131), the Natural Science Basic Research Plan in Shaanxi Province of China (Program No. 2014JQ2-6030), the Scientific Research Program Funded by Shaanxi Provincial Education Department (Program No. 15JK1316), PAPD and CICAEET.


\end{document}